\documentclass{an}
\usepackage{graphicx}
\usepackage{times}
\begin{document}

\Pagespan{1}{}
\Yearpublication{\ldots}
\Yearsubmission{\ldots}
\Month{\ldots}
\Volume{\ldots}
\Issue{\ldots}
\DOI{This.is/not.aDOI}

\title{Photometric monitoring of the blazar 3C~345 for the period 1996 -- 2006\thanks{Based
on observations obtained with the 2-m and 50/70-cm telescopes of the Rozhen National
Astronomical Observatory, and the 60-cm telescope of the Belogradchik
Astronomical Observatory, which are operated by the Institute of Astronomy,
Bulgarian Academy of Sciences, and with the 1.3-m telescope of the Skinakas
Observatory, Crete, Greece; Skinakas Observatory is a collaborative project
of the University of Crete, the Foundation for Research and Technology~--
Hellas, and the Max-Planck-Institut f\"{u}r Extraterrestrische Physik.}}

\author{B. Mihov\thanks{Corresponding author: \email{bmihov@astro.bas.bg}} \and 
R. Bachev \and L. Slavcheva-Mihova \and A. Strigachev \and E. Semkov \and G. Petrov}

\titlerunning{Photometric monitoring of the blazar 3C~345}
\authorrunning{B. Mihov et~al.}

\institute{Institute of Astronomy, Bulgarian Academy of Sciences, 72 Tsarigradsko Chausse
Blvd., 1784 Sofia, Bulgaria}

\received{\ldots}
\accepted{\ldots}
\publonline{later}


\abstract{We present the results of the blazar 3C~345 monitoring
in Johnson-Cousins $BVRI$ bands for the period 1996 -- 2006. We have
collected 29 $V$ and 43 $R$ data points for this period; the $BI$ light
curves contain a few measurements only. The accuracy of our photometry
is not better than 0.03 mag in the $VR$ bands. The total amplitude of the
variability obtained from our data is 2.06 mag in the $V$ band and 2.25
mag in the $R$ one. 3C~345 showed periods of flaring activity during 1998/99
and 2001: a maximum of the blazar brightness was detected in 2001 February~--
15.345 mag in the $V$ band and 14.944 mag in the $R$ one. We confirm that
during brighter stages 3C~345 becomes redder; for higher fluxes the colour
index seems to be less dependent on the magnitude. The intra-night monitoring
of 3C~345 in three consecutive nights in 2001 August revealed no significant
intra-night variability; 3C~345 did not show evident flux changes over
timescales of weeks around the period of the intra-night monitoring. This result
supports the existing facts that intra-night variability is correlated
with rapid flux changes rather than with specific flux levels.}

\maketitle

\section{Introduction}
\label{intro}

Blazars are a sub-class of active galactic nuclei (see the
review paper of Angel \& Stockman \cite{angel80}). The most
notable feature of blazars is their violent variability at
all wavelengths on time scales from about an hour or less
to years. It is now believed that the physical processes in
relativistic jets are responsible for the observed behaviour
of blazars (e.g. Schramm et~al. \cite{schramm93a}; Wagner et~al.
\cite{wagner95}; Otterbein et~al. \cite{otterbein98}; Lobanov \&
Roland \cite{lobanov05}).

The violent variability of blazars is very helpful in
understanding their nature; so, they were targets of a number
of monitoring campaigns like Hamburg Quasar Monitoring (HQM;
Borgeest \& Schramm \cite{borgeest94}; Schramm et~al. \cite{schramm94a},
\cite{schramm94b}) in the past and Whole Earth Blazar Telescope (WEBT)
in the present days. In particular, the international coordinated
programme WEBT proved to be very effective in obtaining blazar
light curves of dense temporal coverage (e.g. Villata et~al. \cite{villata06}).

We started to monitor selected blazars with the 2.0-m telescope of the
Rozhen National Astronomical Observatory (NAO), Bulgaria, in
1996 inspired by Dr. K.-J. Schramm and following the international
collaborations MEGAPHOT (Schramm et~al. \cite{schramm93b}) and Joint
Optical Monitoring Programme of Quasars (JOMPQ; Schramm et~al. \cite{schramm94c}).
The blazar 3C~345 (1641+399, $z=0.5928$) is among the most highly
variable blazars in our list. Regular photometric monitoring of 3C~345
has been carried out since 1965; the historical light curve of the source
was constructed and studied by Schramm et~al. (\cite{schramm93a}), Zhang
et~al. (\cite{zhang98}) and Howard et~al. (\cite{howard04}, hereafter H04).
We present in this paper the results of our monitoring of the blazar 3C~345
for the period 1996 -- 2006.

Another aim of this study is to present our results of the intra-night
monitoring of 3C~345; note that the characteristics of 3C~345 intra-night
variability are not well established. Kidger (\cite{kidger89}) was the first
one who drew attention to the intra-night variability of 3C~345: he detected
flickering on time scales of hours with amplitudes of about 0.1~-- 0.2 mag.
Furthermore, Kidger \& de~Diego (\cite{kidger90}) reported 0.47 $B$ mag
brightness drop in 13 minutes, whereas H04 detected no significant intra-night
variability; H04 found that the occurrences of intra-night variability are
correlated temporally with long-term optical activity of the objects studied.
Thus, intra-night monitoring of 3C~345 was undertaken by us in order to shed
more light on the intra-night variability characteristics of this source.

The paper is organized as follows. The observations and data
reduction are described in Sect.~\ref{obsred}. The photometry and resulting
light curves are presented in Sect.~\ref{photolc}. The results are discussed
in Sect.~\ref{disc}. A brief summary of our results is presented in Sect.~\ref{summ}.

\begin{table}[t]
\caption{Johnson-Cousins $BVRI$ magnitudes of stars \#4 and \#19 in the field of 3C~345.}
\label{refs}
\centering
\begin{tabular}{@{}rrrrr@{}}
\noalign{\smallskip} \hline
\noalign{\smallskip}
Star & $B$ & $V$ & $R$ & $I$ \\
     & $\sigma_B$ & $\sigma_V$ & $\sigma_R$ & $\sigma_I$ \\
\noalign{\smallskip} \hline
\noalign{\smallskip}
\#4  & 16.044 & 15.245 & 14.768 & 14.337 \\
     &  0.017 &  0.007 &  0.006 &  0.011 \\
\noalign{\smallskip} \hline
\noalign{\smallskip}
\#19 & 16.452 & 15.228 & 14.470 & 13.806 \\
     &  0.016 &  0.006 &  0.006 &  0.014 \\
\noalign{\smallskip} \hline
\noalign{\smallskip}
\end{tabular}
\end{table}

\begin{figure}[t]
\resizebox{\hsize}{!}{\includegraphics{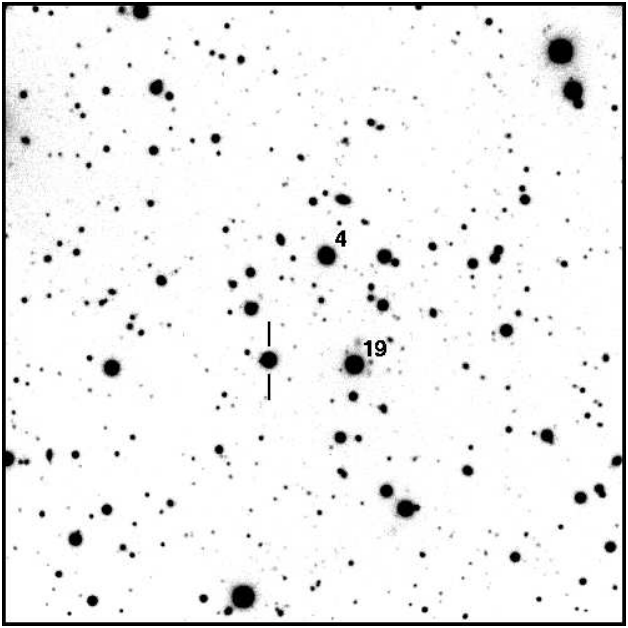}}
\caption{\label{field_r}The field ($8.3 \times 8.3$ arcmin wide) containing
the blazar 3C~345, reference star \#19, and control star \#4. This
image was taken by AS at SO through the $R$ filter; East is to the left,
North is at the top.}
\end{figure}

\section{Observations and data reduction}
\label{obsred}

The observational data of 3C~345 were obtained using the 2.0-m
Ritchey-Chr\'etien and the 0.5/0.7-m Schmidt telescopes of NAO, the
0.6-m Cassegrain telescope of the Belogradchik Astronomical Observatory
(BAO), Bulgaria, and the 1.3-m Ritchey-Chr\'etien telescope of the
Skinakas Observatory (SO), Crete, Greece. Standard Johnson-Cousins
$BVRI$ filters were used in all observations. Focal reducers were occasionally
used at NAO (FoReRo) and BAO. The following CCD cameras were used as detectors
of the 2.0-m telescope of NAO: $375 \times 242$ SBIG ST-6, $1024 \times 1024$
Photometrics AT200, and $1340 \times 1300$ Princeton Instruments VersArray:1300B.
The CCD cameras ST-6 and VersArray:1300B were used in single nights: 1996 August
12/13 and 2005 March 12/13, respectively. The 0.5/0.7-m telescope of NAO and the
0.6-m telescope of BAO were equipped with identical $1530 \times 1020$ SBIG ST-8
CCD cameras. The 1.3-m telescope of SO was equipped with $1024 \times 1024$
Photometrics CCD camera.

Multiple $VR$ frames of the 3C~345 field were taken in each night
allocated for the monitoring; $BI$ frames were taken occasionally.
Twilight flat field, zero exposure, and dark current frames were taken
as well. Dark frames were taken when ST-6/8 CCD cameras were used and
zero frames were taken in the case of the other cameras. The binning
factor of the CCD cameras was changed depending on the observing conditions.

The intra-night monitoring of 3C~345 was performed at BAO by RB during
three nights of 2001 August: 18/19, 19/20, and 20/21. The blazar was
imaged through $VRI$ filters for a period of about 3~-- 4 hours each
night; the exposures were 120 sec for all passbands.

Reduction of the 3C~345 frames was done depending on the CCD camera
used: ST-6/8 CCD data were dark subtracted and flat fielded using the
camera software, whereas the data acquired by means of the other cameras
were de-biased and flat fielded using ESO-MIDAS package. The cosmic ray
hits were cleaned and the individual frames were aligned and co-added
using ESO-MIDAS. The frames obtained in the course of the intra-night
monitoring were dark subtracted and flat fielded only.

\section{Photometry and light curves}
\label{photolc}

We used differential photometry technique to obtain the 3C~345 light curves
in order to be independent of the photometric conditions. Field stars
\#4 and \#19, calibrated by Gonz\'alez-P\'erez, Kidger \& Mart\'in-Luis
(\cite{gonzalez01}), were used as a control star and as a reference one,
respectively (see Table~\ref{refs} and Fig.~\ref{field_r}). Stars \#4 and
\#19 are designated as D and E, respectively, in Smith's et~al. (\cite{smith85})
paper.

The flux measurements of all objects of interest were performed using DAOPHOT
package (Stetson \cite{stetson87}) run within ESO-MIDAS. Instrumental magnitudes were
measured through a set of apertures with radii of $1/2/3 \times \mathrm {FWHM}$
pixels; the sky background value was estimated in a centred annulus with an
inner radius of $7 \times \mathrm {FWHM}$ pixels and containing 1000 pixels.
The calibrated magnitudes of 3C~345 were calculated relative to star \#19 without
taking into account the colour term in the transformation equations; the light
curves of star \#4 were obtained in the same way as the blazar ones and were used
to estimate the accuracy of the photometry.

We adopted as final magnitudes the ones measured through $1 \times \mathrm {FWHM}$
radius aperture~-- using this aperture the control star light curves show the
smallest clipped standard deviation in all passbands. The $3\sigma$ clipping technique
was used to eliminate the deviant data points~-- if such are present~-- in the
control star light curve under the assumption of the control and reference star
non-variability. The blazar data points corresponding to the eliminated control
star points were also eliminated. We got a total of three $V$ band and three $R$ band
data points removed from the light curves by this technique. The formal errors of the
final blazar and control star magnitudes include the errors of the instrumental magnitudes
as returned by DAOPHOT and the errors of the standard magnitudes of reference star \#19.

\begin{figure}[t]
\resizebox{\hsize}{!}{\includegraphics{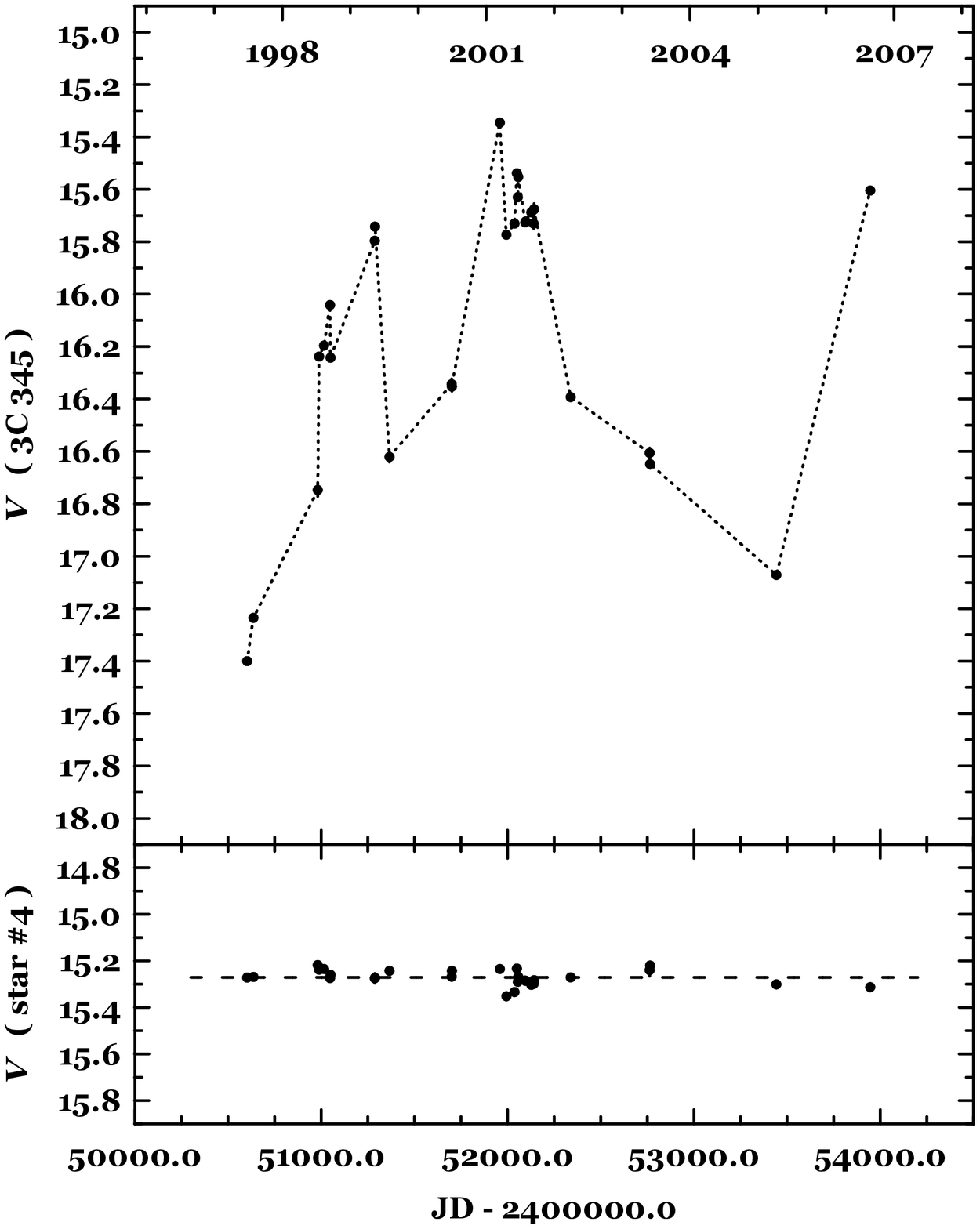}}
\caption{\label{lc_v}$V$ band light curves of the blazar 3C~345 and
of control star \#4. The dashed line represents the standard $V$ band magnitude
of star \#4 (see Table~\ref{refs}). For most of the data points the error
bars are smaller than the symbols.}
\end{figure}

\begin{figure}[t]
\resizebox{\hsize}{!}{\includegraphics{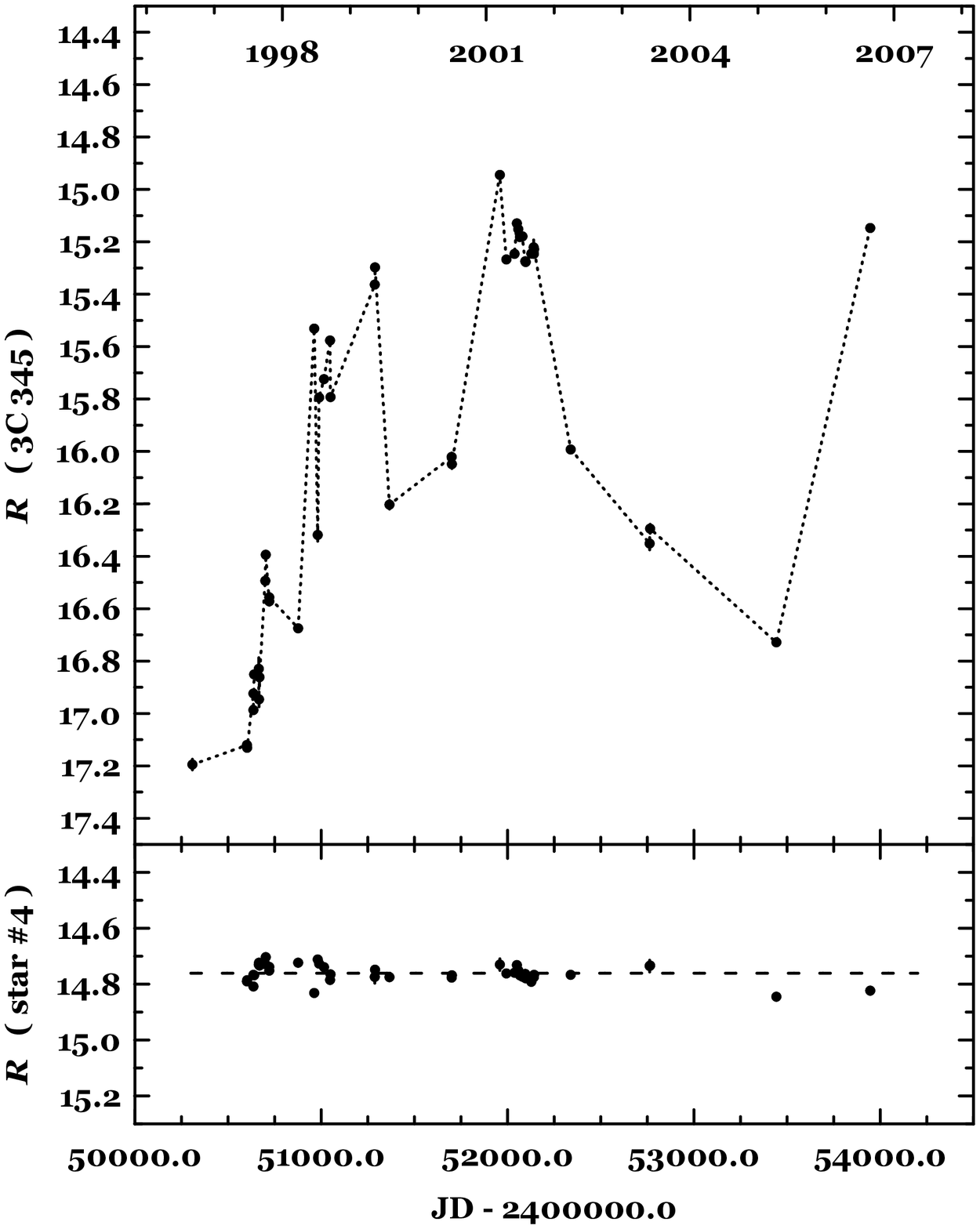}}
\caption{\label{lc_r}The same as in Fig.~\ref{lc_v}, but for the $R$ band.}
\end{figure}

\begin{figure}[t]
\resizebox{\hsize}{!}{\includegraphics{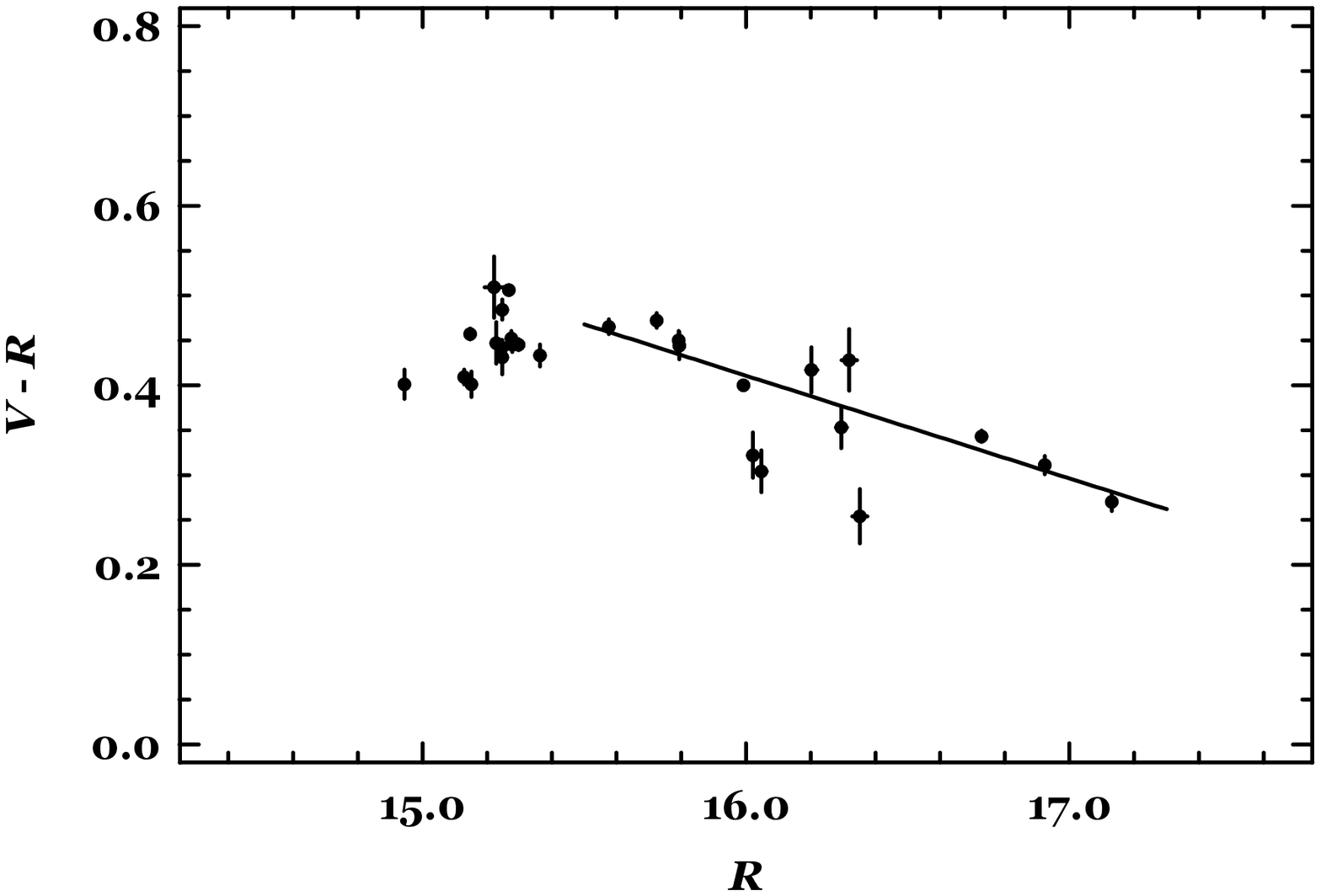}}
\caption{\label{r_vr}Colour index $V-R$ plotted against the $R$ magnitude.
One can see that 3C~345 is redder when it is brighter; a weighted
linear fit is overplotted. Note that for higher fluxes the colour index
seems to be less dependent on the $R$ magnitude.}
\end{figure}

\begin{figure}[t]
\resizebox{\hsize}{!}{\includegraphics{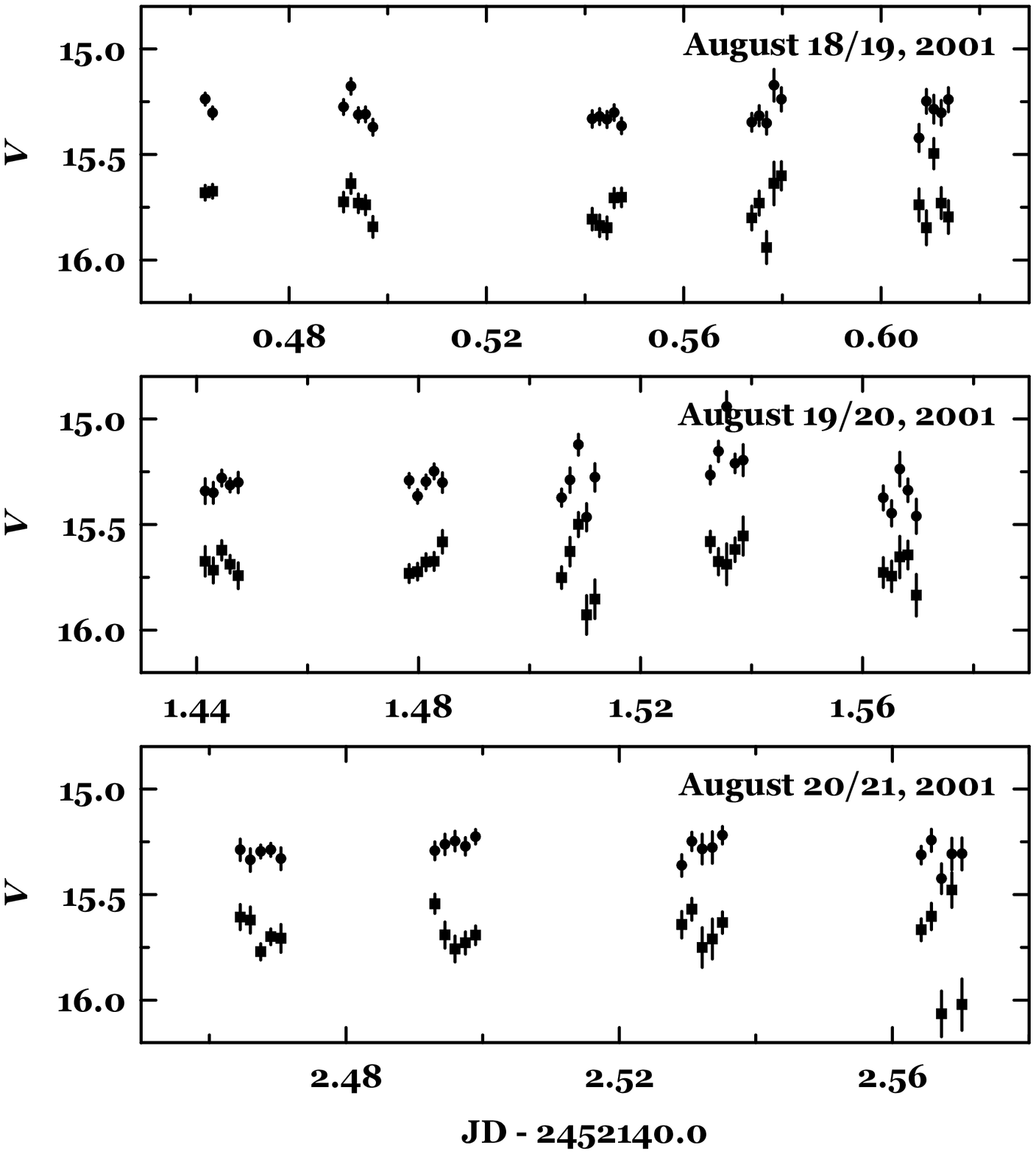}}
\caption{\label{short_v}Intra-night $V$ light curves for three consecutive
nights in 2001 August. Filled squares are the blazar magnitudes and filled
circles~-- star \#4 ones. No significant intra-night variability could
be identified.}
\end{figure}

\begin{figure}[t]
\resizebox{\hsize}{!}{\includegraphics{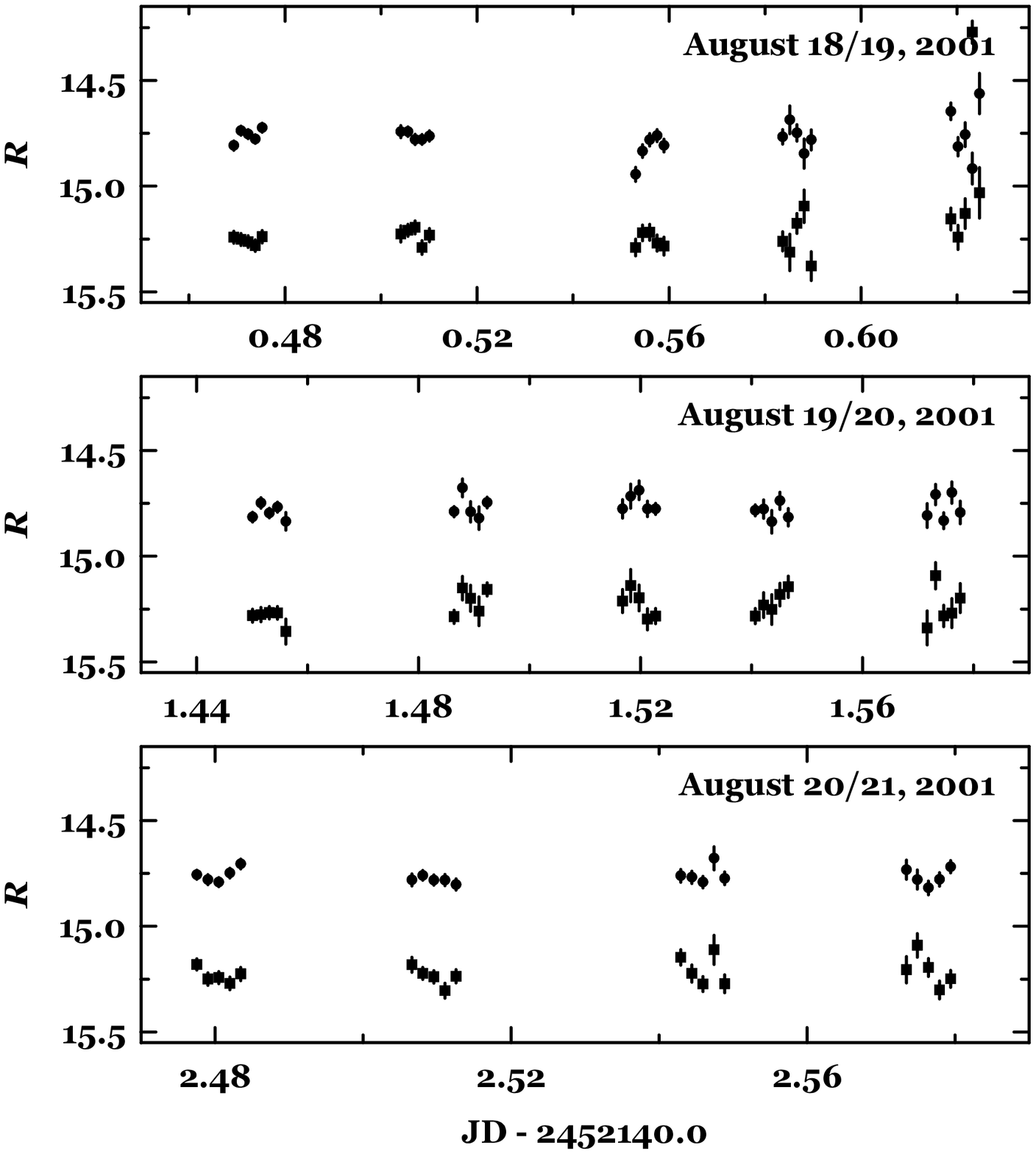}}
\caption{\label{short_r}The same as in Fig.~\ref{short_v}, but for the $R$ band.}
\end{figure}

\begin{figure}[t]
\resizebox{\hsize}{!}{\includegraphics{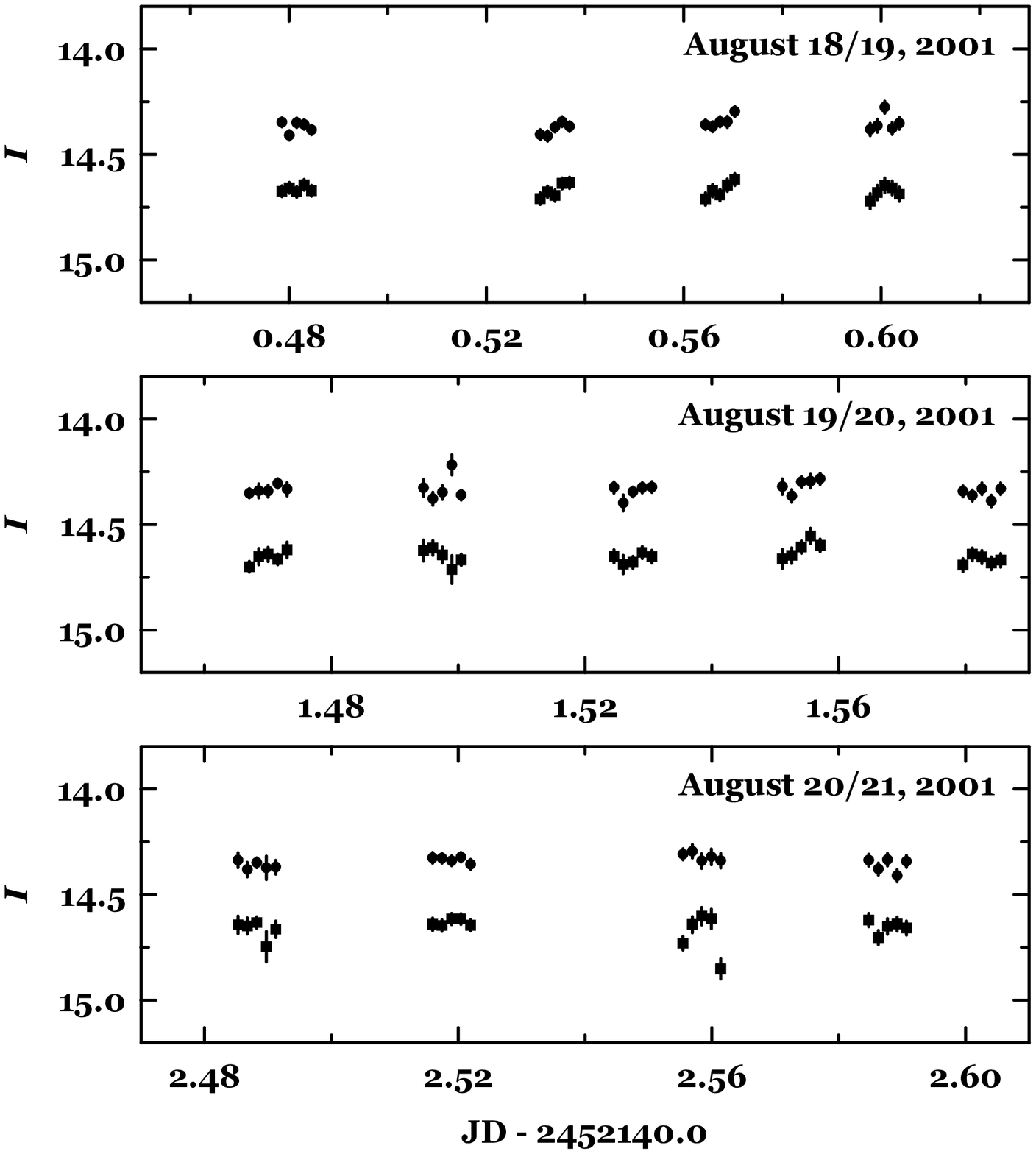}}
\caption{\label{short_i}The same as in Fig.~\ref{short_v}, but for the $I$ band.}
\end{figure}

The nightly weight-mean $VRI$ magnitudes of the blazar were obtained from the
intra-night data and added to the long-term light curve. The errors of the weight-mean
magnitudes were calculated taking the larger between (1) the error estimate
based on the individual magnitude errors, and (2) the error estimate based on the
scatter of magnitudes involved in averaging about their weight-mean value.

The measured light curves of 3C~345 are tabulated in Table~\ref{lc_bt} ($B$ band),
Table~\ref{lc_vt} ($V$ band), Table~\ref{lc_rt} ($R$ band), and Table~\ref{lc_it} ($I$ band)
and are presented in Fig.~\ref{lc_v} ($V$ band) and in Fig.~\ref{lc_r} ($R$ band). The final
light curves contain a total of 4 data points in the $B$ band, 29~-- in the $V$ band,
43~-- in the $R$ band, and 6~-- in the $I$ band. The Universal Time is taken at the middle
of each $(B)VR(I)$ observing set; Julian Days are geocentric. The telescopes
used are abbreviated in a self-explanatory way in the tables. The suffix FR
used in the tables means that focal reducer has been employed. 

Note that due to the small number of $BI$ data points the computed throughout
the paper statistical parameters of the blazar and control star $BI$ light
curves are approximate and should be used with care.

\section{Discussion}
\label{disc}

\subsection{Accuracy of the photometry}
\label{acurr}

The mean $BVRI$ magnitudes of the control star are 16.090, 15.272,
14.761, and 14.355, respectively. These magnitudes are in good agreement
with those presented in Table~\ref{refs}. Consequently, the systematic errors
introduced due to the colour term skipping in magnitude calculation are
rather negligible.

The standard deviations of the control star $BVRI$ light curves are 0.014
mag, 0.033 mag, 0.031 mag, and 0.013 mag, respectively. We found that the standard
deviations of the $VR$ bands are larger than the mean formal errors of the calibrated
3C~345 $VR$ magnitudes by factors of 2.7 and 2.1, respectively. Based on the
scatter of the control star magnitudes, we conclude that the accuracy of our $VR$
photometry is not better than 0.03 mag; actually, the accuracy should be worse
because control star \#4 was brighter than the blazar during the monitoring period.

\subsection{$V-R$ colour index}
\label{colin}

The $V-R$ colour index dependence on the $R$ magnitude is plotted in Fig.~\ref{r_vr}.
Using a weighted linear fit in the interval of the $R$ magnitudes fainter
than $R_0=15.5$ mag, we got the following anti-correlation
$$
V-R=(2.242 \pm 0.226)-(0.114 \pm 0.014)R,
$$
with a correlation coefficient $r=-0.718$, i.e. during the brighter stages
3C~345 becomes redder. The cut-off magnitude $R_0$ was introduced by us as
the dependence of the colour index on the flux seems to be less pronounced
for brighter stages of the source (see Fig.~\ref{r_vr}). The value of $R_0$ was
determined by eye and should be considered as approximate because the region where
the ``colour index -- magnitude" relation changes its slope is not well covered with
data points. Schramm's et~al. (\cite{schramm93a}) Fig.~4 could be regarded as a
support in favour of the cut-off magnitude introduction. However, the presence of a
cut-off magnitude is not so obvious inspecting Zhang's et~al. (\cite{zhang00}) ``colour
index -- magnitude" figures.

Fitting over the entire range of magnitudes, we got
$$
V-R=(1.762 \pm 0.179)-(0.085 \pm 0.011)R,
$$
with a correlation coefficient $r=-0.734$. In this case the fitted coefficients
are in good agreement with those obtained by Schramm et~al. (\cite{schramm93a}) who
used 1991/92 data and by Zhang et~al. (\cite{zhang00}) who used 1991/92 and 1996/97 data.

\subsection{Intra-night variability}
\label{inv}

The $VRI$ intra-night monitoring light curves are presented in Fig.~\ref{short_v},
Fig.~\ref{short_r}, and Fig.~\ref{short_i}, respectively. No significant intra-night
variability patterns or gradients could be identified by eye. To quantify the intra-night
variability of 3C~345 we used the variability parameter $C$ proposed by Jang \& Miller
(\cite{jang97}). Given the standard deviations of the blazar and of the control star
light curves, $\sigma(3\mathrm {C}~345)$ and $\sigma(\mathrm {star}~\#4)$, respectively,
the variability parameter is defined as
$$
C={\sigma(3\mathrm {C}~345) \over \sigma(\mathrm {star}~\#4)}.
$$

If $C > C_0$ $(C_0=2.576)$, then the source is considered variable
at the 99\% confidence level. Note that for the period of the
intra-night monitoring the blazar and the control star were of compatible
brightness; so, the parameter $C$ cannot be a subject to significant systematic
errors due to the different flux densities of the objects considered. A further
complication of the above simple criterion for variability could be considered
when multi-colour observations are available: one could expect correlated
variations in different passbands to be observed, i.e. the median value of $C$
over all passbands for a given night to be larger than $C_0$. The criterion
$C > C_0$ was met in the first night ($C=2.7$ in the $R$ band) and in
the third night ($C=2.9$ in the $V$ band) of the intra-night monitoring. However,
the computed median values of $C$ over the $VRI$ bands are 1.6, 1.0, and 2.1 for
the consecutive nights, respectively. So, we could conclude that for the
period of the intra-night monitoring 3C~345 did not show significant
intra-night variations. 

\begin{figure}[t]
\resizebox{\hsize}{!}{\includegraphics{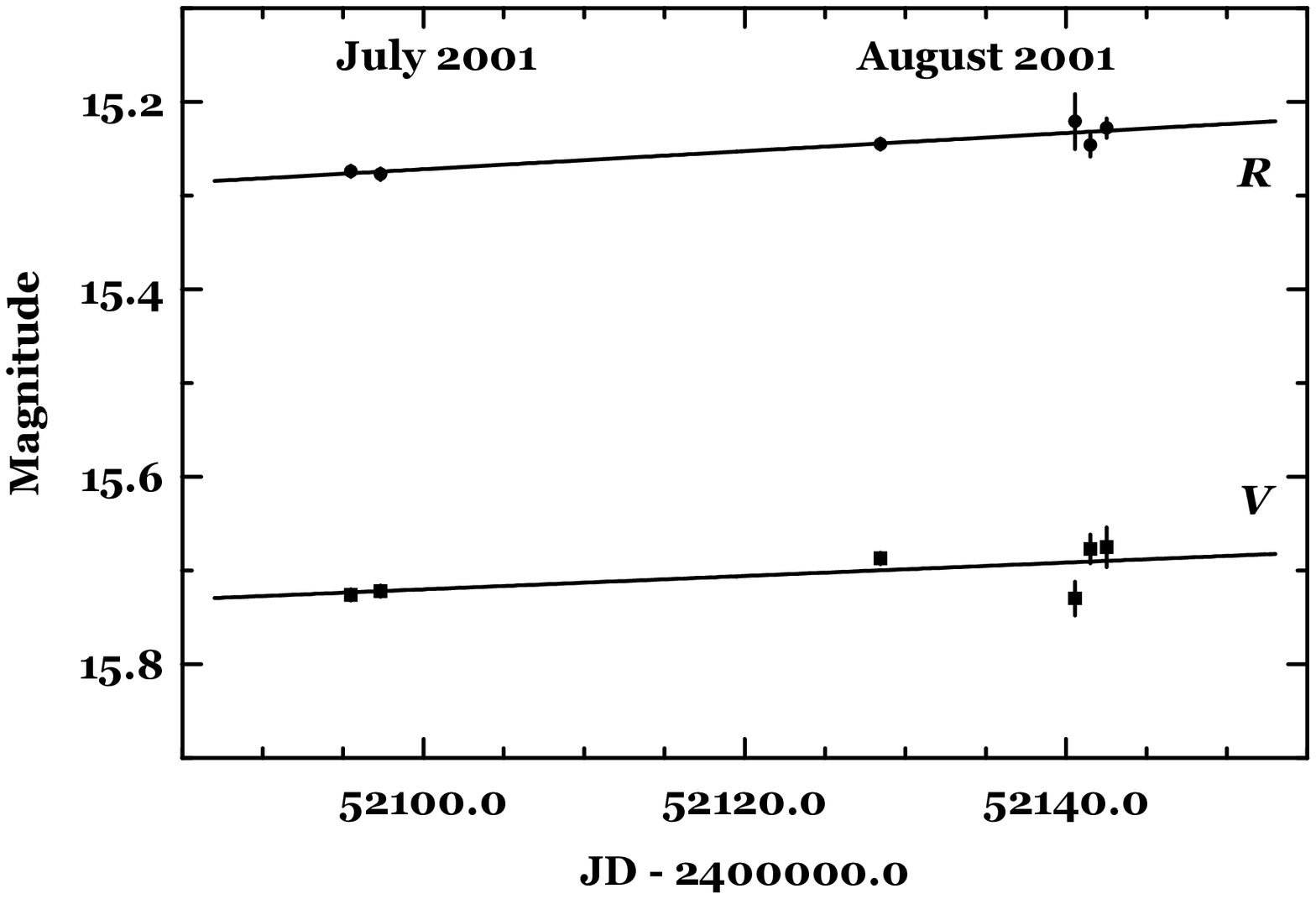}}
\caption{\label{grad}$VR$ light curves of 3C~345 for the period 2001 July/August.
The weighted linear fits used to estimate the light curve gradients are overplotted.}
\end{figure}

The gradients of the $VR$ long-term light curves for the period 2001 July/August,
i.e. around the dates of the intra-night monitoring, are
$-0.32 \pm 0.07$ and $-0.29 \pm 0.06$ mag per year, respectively; the
weight-mean magnitudes computed for the nights of intra-night monitoring were
also included in the gradient calculation. The gradients were estimated using
a weighted linear fit to the 2001 July/August $VR$ data points; this part
of the light curves and the corresponding fits are presented in Fig.~\ref{grad}.
The standard deviations of the $VR$ long-term light curves for the period 2001 July/August
are 0.026 and 0.023 mag, respectively; the same data points were used as in the
gradient calculation. Therefore, 3C~345 did not show significant flux changes over
timescales of weeks around the period of the intra-night monitoring\footnote{One
could see that 2001 July/August part of the long-term light curves is not well sampled;
so, our conclusion should be used with some caution: its truthfulness depends on the
blazar behaviour during the gaps in our monitoring.}. This result is in agreement with H04's
conclusion that the presence of intra-night variability occurs more frequently while the
flux is changing. We observed no evident intra-night variability while the
source was in a bright stage. This supports H04's finding of no correlation between
the presence/absence of intra-night variability and the flux level of sources;
the intra-night variability observations used by H04 were performed when 3C~345
was fainter compared to the period of our intra-night observations.

\subsection{Long-term variability}
\label{long}

One could see from Fig.~\ref{lc_v} and Fig.~\ref{lc_r} that 3C~345 was in a bright
stage during 1998/99 and 2001, i.e. the blazar showed periods of flaring activity.
In particular, a maximum in 3C~345 brightness was detected in 2001 February:
15.345 mag in the $V$ band and 14.944 mag in the $R$ one~-- values compatible
with the 1991/92 outburst ones (see Schramm et~al. \cite{schramm93a}).
Another flare of brightness could be seen at the end of July 2006. Unfortunately,
our measurements are too sparse to be able to follow the individual flares accurately.

The total amplitude of variability detected by us for the period 1996~-- 2006
is 1.40 mag in the $B$ band, 2.06 mag in the $V$ band, 2.25 mag in the $R$ band
and 1.00 mag in the $I$ band; one should keep in mind the very different
sampling of $BI$ and $VR$ light curves.

The high brightness level observed during 2001 is in agreement with Zhang's et~al.
(\cite{zhang98}) prediction that the following large outburst of 3C~345 should be at
its maximum around 2002 January. This prediction was made on the basis of the period
of $10.1 \pm 0.8$ years found by them.

\begin{table*}[t]
\caption{$B$ band light curves of the blazar 3C~345 and of control star \#4.}
\label{lc_bt}
\centering
\begin{tabular}{@{}lrrrrr@{}}
\noalign{\smallskip} \hline
\noalign{\smallskip}
Civil date &    UT & JD $-$ 2400000 &       $B$ (3C~345) &     $B$ (star \#4) & Telescope \\
\noalign{\smallskip} \hline
\noalign{\smallskip}
1997 Jul 05 & 22:39 & 50635.4438 & $17.327 \pm 0.017$ & $16.077 \pm 0.017$ & ROZ2.0 \\
2001 May 07 & 22:36 & 52037.4417 & $16.075 \pm 0.018$ & $16.096 \pm 0.018$ & SKI1.3 \\
2001 May 08 & 00:18 & 52037.5122 & $16.078 \pm 0.017$ & $16.106 \pm 0.017$ & SKI1.3 \\
2006 Jul 29 & 19:14 & 53946.3012 & $15.928 \pm 0.013$ & $16.080 \pm 0.017$ & ROZ2.0 \\
\noalign{\smallskip} \hline
\noalign{\smallskip}
\end{tabular}
\end{table*}

\begin{table*}[t]
\caption{$V$ band light curves of the blazar 3C~345 and of control star \#4.}
\label{lc_vt}
\centering
\begin{tabular}{@{}lrrrrr@{}}
\noalign{\smallskip} \hline
\noalign{\smallskip}
Civil date &    UT & JD $-$ 2400000 &       $V$ (3C~345) &     $V$ (star \#4) & Telescope \\
\noalign{\smallskip} \hline
\noalign{\smallskip}
1997 Jun 03 & 01:07 & 50602.5464 & $17.401 \pm 0.007$ & $15.272 \pm 0.006$ &     ROZ2.0 \\
1997 Jul 05 & 22:39 & 50635.4438 & $17.235 \pm 0.008$ & $15.269 \pm 0.006$ &     ROZ2.0 \\
1998 Jun 17 & 00:23 & 50981.5162 & $16.747 \pm 0.025$ & $15.218 \pm 0.010$ &     BEL0.6 \\
1998 Jun 23 & 22:59 & 50988.4578 & $16.238 \pm 0.009$ & $15.238 \pm 0.007$ &     ROZ2.0 \\
1998 Jul 20 & 22:00 & 51015.4168 & $16.196 \pm 0.006$ & $15.235 \pm 0.006$ &     ROZ2.0 \\
1998 Aug 20 & 20:31 & 51046.3545 & $16.041 \pm 0.006$ & $15.274 \pm 0.006$ &     ROZ2.0 \\
1998 Aug 23 & 20:05 & 51049.3368 & $16.242 \pm 0.007$ & $15.260 \pm 0.007$ &     ROZ2.0 \\
1999 Apr 19 & 00:27 & 51287.5188 & $15.796 \pm 0.009$ & $15.275 \pm 0.020$ &     ROZ2.0 \\
1999 Apr 20 & 01:54 & 51288.5795 & $15.742 \pm 0.004$ & $15.273 \pm 0.004$ &     ROZ2.0 \\
1999 Jul 06 & 22:52 & 51366.4527 & $16.620 \pm 0.019$ & $15.243 \pm 0.009$ &     BEL0.6 \\
2000 Jun 03 & 22:05 & 51699.4201 & $16.343 \pm 0.020$ & $15.268 \pm 0.012$ &   BEL0.6FR \\
2000 Jun 04 & 21:50 & 51700.4097 & $16.352 \pm 0.017$ & $15.243 \pm 0.010$ &   BEL0.6FR \\
2001 Feb 17 & 01:53 & 51957.5784 & $15.345 \pm 0.009$ & $15.235 \pm 0.010$ & ROZ0.5/0.7 \\
2001 Mar 24 & 22:22 & 51993.4318 & $15.773 \pm 0.003$ & $15.352 \pm 0.003$ &   ROZ2.0FR \\
2001 May 07 & 22:36 & 52037.4417 & $15.730 \pm 0.008$ & $15.334 \pm 0.008$ &     SKI1.3 \\
2001 May 20 & 01:16 & 52049.5528 & $15.538 \pm 0.006$ & $15.233 \pm 0.006$ &     SKI1.3 \\
2001 May 25 & 00:16 & 52054.5115 & $15.630 \pm 0.006$ & $15.290 \pm 0.006$ &     SKI1.3 \\
2001 May 27 & 21:26 & 52057.3933 & $15.552 \pm 0.008$ & $15.269 \pm 0.010$ & ROZ0.5/0.7 \\
2001 Jul 04 & 23:37 & 52095.4844 & $15.726 \pm 0.006$ & $15.285 \pm 0.006$ &     SKI1.3 \\
2001 Jul 06 & 19:49 & 52097.3260 & $15.722 \pm 0.006$ & $15.286 \pm 0.006$ &     SKI1.3 \\
2001 Aug 06 & 22:46 & 52128.4490 & $15.687 \pm 0.006$ & $15.303 \pm 0.006$ &     SKI1.3 \\
2001 Aug 19 & 01:08 & 52140.5469 & $15.730 \pm 0.018$ & $15.299 \pm 0.012$ &     BEL0.6 \\
2001 Aug 20 & 00:22 & 52141.5156 & $15.677 \pm 0.015$ & $15.293 \pm 0.017$ &     BEL0.6 \\
2001 Aug 21 & 00:39 & 52142.5273 & $15.675 \pm 0.021$ & $15.283 \pm 0.010$ &     BEL0.6 \\
2002 Mar 05 & 23:31 & 52339.4796 & $16.392 \pm 0.003$ & $15.271 \pm 0.003$ &   ROZ2.0FR \\
2003 May 02 & 21:12 & 52762.3837 & $16.606 \pm 0.019$ & $15.240 \pm 0.025$ & ROZ0.5/0.7 \\
2003 May 05 & 21:16 & 52765.3858 & $16.648 \pm 0.015$ & $15.220 \pm 0.011$ & ROZ0.5/0.7 \\
2005 Mar 12 & 23:58 & 53442.4989 & $17.072 \pm 0.004$ & $15.301 \pm 0.003$ &     ROZ2.0 \\
2006 Jul 29 & 19:14 & 53946.3012 & $15.604 \pm 0.005$ & $15.313 \pm 0.005$ &     ROZ2.0 \\
\noalign{\smallskip} \hline
\noalign{\smallskip}
\end{tabular}
\end{table*}

\begin{table*}[t]
\caption{$R$ band light curves of the blazar 3C~345 and of control star \#4.}
\label{lc_rt}
\centering
\begin{tabular}{@{}lrrrrr@{}}
\noalign{\smallskip} \hline
\noalign{\smallskip}
Civil date &    UT & JD $-$ 2400000 &       $R$ (3C~345) &     $R$ (star \#4) & Telescope \\
\noalign{\smallskip} \hline
\noalign{\smallskip}
1996 Aug 12 & 23:08 & 50308.4637 & $17.195 \pm 0.020$ & $          \ldots$ &     ROZ2.0 \\
1997 Jun 01 & 22:48 & 50601.4500 & $17.121 \pm 0.007$ & $14.790 \pm 0.006$ &     ROZ2.0 \\
1997 Jun 03 & 01:07 & 50602.5464 & $17.131 \pm 0.007$ & $14.786 \pm 0.006$ &     ROZ2.0 \\
1997 Jul 05 & 22:39 & 50635.4438 & $16.924 \pm 0.006$ & $14.808 \pm 0.006$ &     ROZ2.0 \\
1997 Jul 06 & 21:27 & 50636.3934 & $16.987 \pm 0.007$ & $14.767 \pm 0.006$ &     ROZ2.0 \\
1997 Jul 10 & 19:17 & 50640.3032 & $16.851 \pm 0.007$ & $14.768 \pm 0.006$ &     ROZ2.0 \\
1997 Aug 03 & 23:14 & 50664.4680 & $16.830 \pm 0.043$ & $14.730 \pm 0.011$ &     BEL0.6 \\
1997 Aug 04 & 21:31 & 50665.3965 & $16.946 \pm 0.028$ & $14.723 \pm 0.008$ &     BEL0.6 \\
1997 Aug 07 & 22:44 & 50668.4474 & $16.862 \pm 0.033$ & $14.733 \pm 0.009$ &     BEL0.6 \\
1997 Sep 07 & 19:42 & 50699.3210 & $16.494 \pm 0.006$ & $14.728 \pm 0.006$ &     ROZ2.0 \\
1997 Sep 10 & 19:42 & 50702.3212 & $16.394 \pm 0.009$ & $14.703 \pm 0.006$ &     ROZ2.0 \\
1997 Sep 28 & 18:10 & 50720.2567 & $16.572 \pm 0.014$ & $14.738 \pm 0.007$ &     BEL0.6 \\
1997 Sep 29 & 18:27 & 50721.2690 & $16.557 \pm 0.016$ & $14.752 \pm 0.007$ &     BEL0.6 \\
1998 Mar 04 & 01:09 & 50876.5479 & $16.675 \pm 0.006$ & $14.723 \pm 0.006$ &     ROZ2.0 \\
1998 May 28 & 23:43 & 50962.4882 & $15.531 \pm 0.010$ & $14.832 \pm 0.008$ &   BEL0.6FR \\
1998 Jun 17 & 00:23 & 50981.5162 & $16.319 \pm 0.023$ & $14.712 \pm 0.010$ &     BEL0.6 \\
1998 Jun 23 & 22:59 & 50988.4578 & $15.794 \pm 0.012$ & $14.727 \pm 0.009$ &     ROZ2.0 \\
1998 Jul 20 & 22:00 & 51015.4168 & $15.724 \pm 0.006$ & $14.739 \pm 0.006$ &     ROZ2.0 \\
1998 Aug 20 & 20:31 & 51046.3545 & $15.576 \pm 0.006$ & $14.785 \pm 0.006$ &     ROZ2.0 \\
1998 Aug 23 & 20:05 & 51049.3368 & $15.792 \pm 0.007$ & $14.765 \pm 0.006$ &     ROZ2.0 \\
1999 Apr 19 & 00:27 & 51287.5188 & $15.363 \pm 0.008$ & $14.774 \pm 0.022$ &     ROZ2.0 \\
1999 Apr 20 & 01:54 & 51288.5795 & $15.297 \pm 0.004$ & $14.748 \pm 0.005$ &     ROZ2.0 \\
1999 Jul 06 & 22:52 & 51366.4527 & $16.203 \pm 0.017$ & $14.775 \pm 0.009$ &     BEL0.6 \\
2000 Jun 03 & 22:05 & 51699.4201 & $16.021 \pm 0.015$ & $14.777 \pm 0.009$ &   BEL0.6FR \\
2000 Jun 04 & 21:50 & 51700.4097 & $16.048 \pm 0.016$ & $14.768 \pm 0.009$ &   BEL0.6FR \\
2001 Feb 17 & 01:53 & 51957.5784 & $14.944 \pm 0.013$ & $14.730 \pm 0.020$ & ROZ0.5/0.7 \\
2001 Mar 24 & 22:22 & 51993.4318 & $15.267 \pm 0.003$ & $14.762 \pm 0.003$ &   ROZ2.0FR \\
2001 May 07 & 22:36 & 52037.4417 & $15.246 \pm 0.008$ & $14.759 \pm 0.007$ &     SKI1.3 \\
2001 May 20 & 01:16 & 52049.5528 & $15.129 \pm 0.006$ & $14.732 \pm 0.006$ &     SKI1.3 \\
2001 May 27 & 21:26 & 52057.3933 & $15.151 \pm 0.011$ & $14.751 \pm 0.006$ & ROZ0.5/0.7 \\
2001 Jun 05 & 23:27 & 52066.4771 & $15.182 \pm 0.006$ & $14.768 \pm 0.006$ &     SKI1.3 \\
2001 Jun 19 & 20:40 & 52080.3608 & $15.179 \pm 0.006$ & $14.773 \pm 0.006$ &     SKI1.3 \\
2001 Jul 04 & 23:38 & 52095.4844 & $15.274 \pm 0.006$ & $14.763 \pm 0.006$ &     SKI1.3 \\
2001 Jul 06 & 19:50 & 52097.3260 & $15.277 \pm 0.006$ & $14.779 \pm 0.006$ &     SKI1.3 \\
2001 Aug 06 & 22:46 & 52128.4490 & $15.245 \pm 0.006$ & $14.792 \pm 0.006$ &     SKI1.3 \\
2001 Aug 19 & 01:08 & 52140.5469 & $15.221 \pm 0.029$ & $14.769 \pm 0.010$ &     BEL0.6 \\
2001 Aug 20 & 00:22 & 52141.5156 & $15.246 \pm 0.012$ & $14.775 \pm 0.008$ &     BEL0.6 \\
2001 Aug 21 & 00:39 & 52142.5273 & $15.228 \pm 0.010$ & $14.767 \pm 0.007$ &     BEL0.6 \\
2002 Mar 05 & 23:31 & 52339.4796 & $15.992 \pm 0.003$ & $14.767 \pm 0.003$ &   ROZ2.0FR \\
2003 May 02 & 21:12 & 52762.3837 & $16.352 \pm 0.023$ & $14.735 \pm 0.021$ & ROZ0.5/0.7 \\
2003 May 05 & 21:16 & 52765.3858 & $16.295 \pm 0.017$ & $14.733 \pm 0.013$ & ROZ0.5/0.7 \\
2005 Mar 12 & 23:58 & 53442.4989 & $16.729 \pm 0.004$ & $14.845 \pm 0.003$ &     ROZ2.0 \\
2006 Jul 29 & 19:14 & 53946.3012 & $15.147 \pm 0.004$ & $14.823 \pm 0.009$ &     ROZ2.0 \\
\noalign{\smallskip} \hline
\noalign{\smallskip}
\end{tabular}
\end{table*}

\begin{table*}[t]
\caption{$I$ band light curves of the blazar 3C~345 and of control star \#4.}
\label{lc_it}
\centering
\begin{tabular}{@{}lrrrrr@{}}
\noalign{\smallskip} \hline
\noalign{\smallskip}
Civil date &    UT & JD $-$ 2400000 &       $I$ (3C~345) &     $I$ (star \#4) & Telescope \\
\noalign{\smallskip} \hline
\noalign{\smallskip}
2000 Jun 04 & 21:50 & 51700.4097 & $15.548 \pm 0.018$ & $14.364 \pm 0.015$& BEL0.6FR \\
2001 May 07 & 22:36 & 52037.4417 & $14.704 \pm 0.015$ & $14.361 \pm 0.015$&   SKI1.3 \\
2001 Aug 19 & 01:08 & 52140.5469 & $14.668 \pm 0.006$ & $14.361 \pm 0.007$&   BEL0.6 \\
2001 Aug 20 & 00:22 & 52141.5156 & $14.650 \pm 0.007$ & $14.335 \pm 0.006$&   BEL0.6 \\
2001 Aug 21 & 00:39 & 52142.5273 & $14.650 \pm 0.010$ & $14.343 \pm 0.006$&   BEL0.6 \\
2006 Jul 29 & 19:14 & 53946.3012 & $14.547 \pm 0.010$ & $14.368 \pm 0.010$&   ROZ2.0 \\
\noalign{\smallskip} \hline
\noalign{\smallskip}
\end{tabular}
\end{table*}

\section{Summary}
\label{summ}

We have presented the results of the blazar 3C~345 monitoring in Johnson-Cousins
$BVRI$ bands for the period 1996 -- 2006. The total amplitude of variability obtained
out of our data is 2.06 mag in the $V$ band and 2.25 mag in the $R$ one. 3C~345 showed
periods of flaring activity during 1998/99 and 2001: a maximum of the blazar brightness
was detected in 2001 February~-- 15.345 mag in the $V$ band and 14.944 mag in the $R$
one. The intra-night monitoring of 3C~345 in three consecutive nights in 2001 August
did not reveal significant intra-night variability.

Our measurements should be considered as a part of the international efforts aimed
to obtain a dense temporal coverage of the light curve of 3C~345. The availability
of well sampled multi-frequency light curves is of importance to reveal the
source of the blazar activity (e.g. Schramm et~al. \cite{schramm93a};
Lobanov \& Roland \cite{lobanov05}).

The tabulated long-term $BVRI$ and intra-night $VRI$ light curves of 3C~345
and of star \#4 could be found at {\em www.astro.bas.bg/${\sim}$bmihov}.

\acknowledgements
The authors are thankful to the anonymous referee whose constructive suggestions
and criticism helped us to improve this paper.

The authors are thankful to Prof. Y. Papamastorakis and I. Papadakis
for the telescope time at Skinakas Observatory.

The European Southern Observatory Munich Image Data Analysis System (ESO-MIDAS)
is developed and maintained by the European Southern Observatory.

The design and manufacturing of the the Focal Reducer Rozhen (FoReRo)
were performed in the workshop of the Institute of Astronomy, Bulgarian
Academy of Sciences, with financial support by the Ministry of Education
and Science, Bulgaria (contract F-482/2201).

The SBIG ST-8 model CCD camera at the Belogradchik Astronomical Observatory
is provided by the Alexander von Humboldt Foundation, Germany.

We also acknowledge the support by UNESCO-ROSTE for the regional collaboration.

\end{document}